\documentclass[12pt,a4paper]{article}
\usepackage{amsmath,amssymb,amsfonts,amsbsy}

\def\p{\partial}

\def\comb#1#2#3{{\mathsurround 0pt\hbox to 0pt {\hspace*{#3}\raisebox{#2}{${#1}$}\hss}}}
\def\combs#1#2#3{{\mathsurround 0pt\hbox to 0pt {\hspace*{#3}\raisebox{#2}{${\scriptstyle #1}$}\hss}}}
\def\combss#1#2#3{{\mathsurround 0pt\hbox to 0pt {\hspace*{#3}\raisebox{#2}{${\scriptscriptstyle #1}$}\hss}}}
\def\e#1{\mathrm{e}^{#1}}
\def\ii{\mathrm{i}}

\def\bFem{\mathbf{F}}
\def\eqdef{\doteqdot}
\def\Fem{F}
\def\baab{\mathbf{b}}
\def\bEem{\mathbf{E}}

\def\bBem{\mathbf{B}}

\def\bhim{\boldsymbol{\imath}}
\def\xp#1{\comb{\cdot}{-0.9ex}{0.3ex}{#1}}
\def\partsol#1{\mathchoice{\combs{\circ}{1.65ex}{0.5ex}{#1}}{\combs{\circ}{1.65ex}{0.5ex}{#1}}{\combss{\circ}{1.2ex}{0.3ex}{#1}}{}{}}
\def\bHCLor{\boldsymbol{\Lambda}}
\def\invop#1{{#1^{\!\!-\!1}}}
\def\metr{\mathfrak{m}}
\def\mass{{\sf m}}
\def\hgamma{\hat{b}}
\def\hunitc{\hat{1}}

\begin{document}
\textwidth=135mm
 \textheight=200mm
\begin{center}
{\LARGE \bfseries Electromagnetic soliton-particle with spin and magnetic moment%
\footnote{{\small Based on talk at the
20th International Symposium on Spin Physics (SPIN2012),
JINR, Dubna, September 17--22, 2012.
}}
}
\vskip 5mm
{\large\bfseries Alexander A. Chernitskii}$^{\dag,\ddag}$
\vskip 5mm
{\small {\it $^\dag$ State University of Engineering and Economics,}}\\
{\small {\it
%\phantom{$^\dag$}
Marata str. 27, St.-Petersburg, Russia, 191002}}\\
{\small {\it $^\ddag$ A. Friedmann Laboratory for Theoretical Physics, St.-Petersburg, Russia}}
\\
\end{center}
\vskip 5mm
{\small
\centerline{\bf Abstract}
Electromagnetic soliton-particle 
with both quasi-static and quick-oscillating wave parts is considered.
Its mass, spin, charge, and magnetic moment appear naturally when the interaction with distant solitons is considered.
The substantiation of Dirac equation for the wave part of the interacting soliton-particle is given.
}
\vskip 10mm
\section{Soliton-particle}

 The specific electromagnetic solitons can be consider as elementary particles.
The various aspects of this subject was discussed in my previous works
(see for example \cite{Chernitskii1999,Chernitskii2004a,Chernitskii2007a}).

The simplest electromagnetic soliton as a particle was
considered by Gustav Mie
(1869 -- 1957) in the beginning of the
twentieth century (1912 -- 1913).
But of course the remarkable properties of solitons were not
known at that time.
Also the term
``soliton''
%<<soliton>>
had appeared much later.
G. Mie considered a nonlinear electromagnetic model
which has not the gauge invariance.
M. Born, L. Infeld, E. Schr\"odinger, and others considered
the gauge invariant model which is called now as Born -- Infeld
electrodynamics. This model is based on geometrical considerations by A.~Eddington and A.~Einstein in the subject of unified field theory \cite{Chernitskii2006c}.

G. Mie and M. Born considered in their works the simplest spherically symmetrical solution of nonlinear electrodynamics as a model of electron.

But of course the spherically symmetrical field configuration can not describe the real electron which has the spin and the magnetic moment.
Moreover in general case the soliton solution can contain the quick-oscillating wave part. This wave part gives the wave properties of the soliton-particle.

\section{Soliton wave-particle and\\ the electromagnetic interaction}

Here we use the hypercomplex representation for electrodynamics \cite{Chernitskii2002a}. Thus we have the bivector of electromagnetic field as
\begin{equation}
\label{72409883}
\bFem \eqdef \frac{1}{2}\,\Fem_{\mu\nu}\,\baab^{\mu}\,\baab^{\nu}
{}={} \bEem  {}+{} \bhim\,\bBem
\;,
\end{equation}
where $\baab^{\mu}$ ($\mu= 0,1,2,3$) are basis vectors which are the elements of Clifford algebra.
%We have noncommutative product in this algebra.
This algebra well known in connection with Dirac matrix theory. It has $16$ basis elements.
We designate the last element of this algebra (Dirac matrix $\gamma^5$) as $\bhim$.
 Its representation for minkowskian metric is the following \cite{Chernitskii2005a}:
$\bhim = - \baab^0\,\baab^1\,\baab^2\,\baab^3$.

Let us consider the electromagnetic soliton which  is time-periodic in proper coordinate system.
For this coordinate system we have
\begin{equation}
\label{35335854}
\xp{\partsol{\bFem}} = \sum_{n=-\infty}^{\infty}\xp{\partsol{\bFem}}_{\!|n}\,\e{-\bhim\,n\,\xp{\omega}\,\xp{x}^{0}}
\;,
\end{equation}
where $\xp{\partsol{\bFem}}$ is electromagnetic soliton field in its proper coordinate
system $\{\xp{x}^\mu\}$,
$\xp{\partsol{\bFem}}_{\!|n} = \xp{\partsol{\bFem}}_{\!|n} (\xp{x}^1,\xp{x}^2,\xp{x}^3)$ depends on space coordinates.

To obtain the traveling soliton solution $\partsol{\bFem}$
we make the Lorentz transformation for the rest soliton (\ref{35335854}) with the following
formula:
\begin{equation}
\label{LorTrHyp1}
\partsol{\bFem} = \bHCLor\, \xp{\partsol{\bFem}}\,\invop{\bHCLor}
\;,
\end{equation}
where  $\bHCLor$ is the appropriate hypercomplex number called biquaternion.

Then according to (\ref{35335854}) and (\ref{LorTrHyp1}) we have the following representation:
\begin{equation}
\label{353358541}
\partsol{\bFem} = \sum_{n=-\infty}^{\infty}\partsol{\bFem}_{\!|n}\,\e{-\bhim\,n\,k_{\mu}\,x^{\mu}}
\;,
\end{equation}
where $\partsol{\bFem}_{\!|n} = \bHCLor\, \xp{\partsol{\bFem}}_{\!|n}\,\invop{\bHCLor}$.

Now we can divide the traveling soliton field $\partsol{\bFem}$  to two part.
The first part $\partsol{\bFem}_{\!|0}$ is quasi-static.
The second part
\begin{equation}
\label{3533585411}
\sum_{n=1}^{\infty}\left(\partsol{\bFem}_{\!|n}\,\e{-\bhim\,n\,k_{\mu}\,x^{\mu}}+\partsol{\bFem}_{\!|-n}\,\e{\bhim\,n\,k_{\mu}\,x^{\mu}}\right)
\end{equation}
is the quick-oscillating wave packet with dispersion relation
\begin{equation}
\label{762431491}
\metr^{\mu\nu}\,k_{\mu}\,k_{\nu}  = \metr^{00}\,\xp{\omega}^{2}
\;.
\end{equation}

  When we investigate the interaction of the quasi-static electromagnetic charged soliton with small external field (of distant solitons)
  we obtain the classical motion equation with Lorentz force \cite{Chernitskii1999}.
  This trajectory equation appears naturally as manifestation of nonlinearity of the field model.
  If the quasi-static electromagnetic soliton has the own full angular momentum and magnetic moment then
  the nonlinear theory gives the appropriate motion equation of classical spinning particle \cite{Chernitskii2006a}.
    In this case the angular momentum has the field nature. It is calculated by integration over space of
    angular momentum density.

 The quick-oscillating part of traveling soliton is connected with its quasi-static part by the appropriate soliton solution.
This connection gives the transformation of dispersion relation for the wave packet (\ref{3533585411})
from (\ref{762431491})
to the following:
\begin{equation}
\label{799896141}
 \metr^{\mu\nu} \left(k_{\mu}+\frac{q\,\tilde{A}_{\mu}}{\hbar}\right)\left(k_{\nu}+\frac{q\,\tilde{A}_{\nu}}{\hbar}\right)  = \metr^{00}\,\xp{\omega}^{2}
\;,
\end{equation}
where $\tilde{A}_{\mu}$ is the electromagnetic potential of distant solitons.
Here $q$ is electrical charge of the soliton-particle, $\hbar$ is the dimension constant such that
\begin{equation}
\label{810556861}
 \hbar\,\xp{\omega}= \mass
\end{equation}
is the full energy of static part of rest soliton.

The conditions (\ref{799896141}) and (\ref{810556861}) provide that the quick-oscillating part do not go away from the quasi-static part.
That is the velocity of the quasi-static part coincides with the group velocity of the soliton wave packet.

  The quick-oscillating and quasi-static parts have the hard connection in the soliton localization domain where the field is high.
    In this domain the quasi-static part drives the quick-oscillating part.

Outside of the soliton localization domain the field is small and the appropriate connection is weak.
    In this domain the quick-oscillating part is almost free.
In this domain the wave packet is diffracting and interfering.
    From this domain the quick-oscillating part drives the quasi-static part.

\section{Dirac equation}

While we do not know the exact soliton solution and the model equations in detail
we can try to investigate the wave behavior of the soliton-particle by the consideration of a linear wave
with the same dispersion relation (\ref{799896141}).

According to \cite{Chernitskii2002a} the nonlinear electrodynamics equations can be represented in like Dirac equation form. Thus this form of equation
must be natural for the linear wave equation mentioned above. In particular the dispersion relation (\ref{799896141})
is realized for the Dirac equation written in form
\begin{equation}
\label{573935891}
 \hgamma^{\mu}\left(\hbar\,\frac{\p}{\p x^{\mu}} + \ii\,q\,\tilde{A}_{\mu}\right)\psi  = \hunitc\,\mass\,\psi
\;,
\end{equation}
where $\hgamma^{\mu}$ are Dirac matrices such as in \cite{Chernitskii2002a}, $\ii$ is customary imaginary unit,
$\hunitc$ is the unit matrix.

After substitution $\psi=\bar{\psi}\,\e{\ii\,k_{\mu}\,x^{\mu}}$ ($\bar{\psi}$ is constant column)  to (\ref{573935891}) the solvability condition of the linear
algebraical system for $\bar{\psi}$
\begin{equation}
\label{778320231}
\det\!\left[\hgamma^{\mu}\,\ii\left(\hbar\,k_{\mu} + q\,\tilde{A}_{\mu}\right)  - \hunitc\,\mass\right]  = 0
\end{equation}
gives the dispersion relation
(\ref{799896141}).

%The 
Connection between electromagnetic field bivector and four-component spinor was discussed in \cite{Chernitskii2002a}.

\end{document}